\documentclass[aps,pra,twocolumn,showpacs,showkeys,10pt,nofootinbib]{revtex4-1}

\usepackage{amsmath,amssymb,upgreek}
\usepackage{graphicx,color}

\graphicspath{ {.} {./pics/} }

\newcommand{\rme}{\mathrm e}
\newcommand{\rmi}{\mathrm i}
\newcommand{\rmd}{\mathrm d}

\renewcommand{\Re}{\mathop{{\rm Re}}}

\begin{document}

\title{Non-Markovian dynamics of few emitters in a laser-driven cavity}

\author{D. Pagel}
\email{pagel@physik.uni-greifswald.de}
\author{H. Fehske}
\affiliation{Institut f\"ur Physik, Ernst-Moritz-Arndt-Universit\"at, 17487 Greifswald, Germany}

\begin{abstract}

We study the laser-driven Dicke model beyond the rotating-wave approximation.
For weak coupling of the system to environmental degrees of freedom the dissipative dynamics of the emitter-cavity system is described by the Floquet master equation.
Projection of the system evolution onto the emitter degrees of freedom results in non-Markovian behavior.
We quantify the non-Markovianity of the resulting emitter dynamics and show that this quantity can be used as an indicator of the dissipative quantum phase transition occurring at high driving amplitudes.

\end{abstract}



\maketitle

\paragraph*{Introduction.}
Understanding the evolution of open quantum systems that are coupled to the environment is a challenging and important task in quantum optics~\cite{car99, bp02, aga13}.
Usually the most general representation of the quantum dynamical semigroup in Lindblad form is used for the theoretical description~\cite{gks76, lin76}.
Then the open system dynamics is governed by a Markovian master equation that accounts for the decay of information from the system to the environment~\cite{alz06, bgb11, sch14}.

In many realistic physical systems however, the Markovian approximation fails to give an accurate picture of the dynamics.
As a consequence, a variety of analytical and numerical approaches have been developed~\cite{pmhs08, wecc08, rhp10, dva17, blpv16, blp09, lpb10}.
Nevertheless, until recently, not even a rigorous definition of quantum non-Markovianity was available which is independent of the specific representation of the system dynamics.
Meanwhile such a definition is given in terms of the trace distance of two quantum states~\cite{blp09, lpb10}.
The idea behind is that a Markovian process always reduces the distinguishability of any two states.
The unique property of non-Markovian dynamics is the increase of the trace distance for finite intervals of time.
In this sense non-Markovianity is closely related to an information flow from the environment back to the open system~\cite{wc08, blp09, lpb10}.

A paradigmatic system for the theoretical modeling of light-matter interaction is the Dicke model of two-level emitters interacting with a cavity photon mode~\cite{dic54}.
It has been used to study spontaneous emission and superradiance~\cite{bsh71, gh82, mh10, berb12}, cooperativity and lasing~\cite{agpds11, lldvfvb11, mgsa13}, and the emission of non-classical light~\cite{at91, que12, rsh13, paf15}.
In the regime of strong emitter-cavity coupling, the standard scenario of photon blockade is modified by two-photon cascade decay significantly changing the statistics of emitted photons~\cite{srdshs13, paf15}.
Including an additional laser-driving of the cavity, even the dynamic Stark effect can be observed in the emission spectrum~\cite{paf17}.
For a single emitter and laser amplitudes larger than the emitter-cavity coupling strength, a dissipative quantum phase transition occurs, where the system undergoes dressed state polarization~\cite{sc88, ac91, car15, agc92}.
The counterpart in the limit of infinitely many emitters is the superradiance transition~\cite{bsh71, gh82, dtdlss13}.

In this Rapid Communication, we apply the concept of non-Markovianity to the laser-driven Dicke model by studying the dynamics of few emitters in the regimes of strong emitter-cavity coupling and high driving amplitudes.
While the overall system dynamics can be described with a Born-Markov master equation, the projection onto the emitter degrees of freedom yields a non-Markovian evolution.
The quantification of the non-Markovianity provides us with a measure for the information flow between emitters and cavity.
We find driving strength independent results over a wide parameter range and show that the non-Markovianity change is an indicator of the dissipative quantum phase transition at high driving amplitudes.

\paragraph*{Laser-driven Dicke system.}
The interaction of $N$ two-level emitters with a single cavity photon mode is modeled by the Dicke Hamiltonian (with $\hbar=1$)~\cite{dic54}
\begin{equation}\label{HD}
  H_D = \omega_c a^\dagger a + \omega_x \sum_{j=1}^N \sigma_+^{(j)} \sigma_-^{(j)} + g (a + a^\dagger) \sum_{j=1}^N \big( \sigma_-^{(j)} + \sigma_+^{(j)} \big) \,.
\end{equation}
Here, the operator $a$ ($a^\dagger$) annihilates (creates) a cavity photon with frequency $\omega_c$.
Relaxation (excitation) of the $j$th emitter with transition energy $\omega_x$ is taken into account by the operator $\sigma_-^{(j)}$ ($\sigma_+^{(j)}$).
The emitter-cavity coupling strength is $g$.

\begin{figure}
  \includegraphics[width=0.9\linewidth]{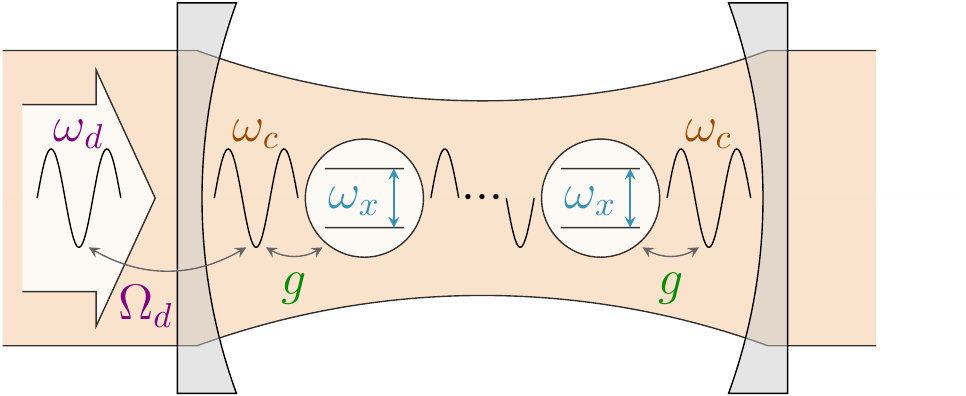}
  \caption{\label{fig:cavity} Two-level emitters in a laser-driven cavity.}
\end{figure}

The Dicke system~\eqref{HD} is driven by an external monochromatic laser of frequency $\omega_d$ (see Fig.~\ref{fig:cavity}).
We consider a driving of the cavity mode (driving of the emitters is obtained from a unitary transformation~\cite{ac91}) which is described by the time-dependent Hamiltonian
\begin{equation}\label{Hd}
  H_d(t) = \Omega (a + a^\dagger) \cos \omega_d t \,,
\end{equation}
where $\Omega$ is the drive amplitude.
Hence, the total system Hamiltonian,
\begin{equation}
  H_S(t) = H_D + H_d(t) \,,
\end{equation}
has a periodic time-dependence, $H_S(t)=H_S(t+T_d)$ with driving period $T_d=2\pi/\omega_d$.
We remark that the counter-rotating interaction terms in the Hamiltonians~\eqref{HD} and~\eqref{Hd} are necessary in the regimes of strong emitter-cavity coupling and high driving amplitudes such that a transformation into a rotating frame is not appropriate.
Instead, we will use the Floquet states~\cite{flo83} $|\psi_n(t)\rangle=\rme^{-\rmi\epsilon_nt}|\phi_n(t)\rangle$ for our analysis of non-Markovian behavior, where $\epsilon_n\in\mathbb{R}$ are the quasienergies and $|\phi_n(t)\rangle=|\phi_n(t+T_d)\rangle$ are periodic wavefunctions.

It is known that a dissipative quantum phase transition (spontaneous dressed state polarization) occurs at $\Omega = g$ if the counter-rotating interaction terms are dropped from the Hamiltonian $H_S(t)$ and only a single emitter is considered~\cite{sc88, ac91, agc92, car15}.
The neoclassical theory presented in Ref.~\cite{car15} can straightforwardly be extended to situations with more than one emitter.
In particular, the semiclassical equations of motion in a frame rotating with the driving frequency (in rotating-wave approximation) are
\begin{eqnarray}\label{dalpha}
  \frac{\rmd}{\rmd t} \alpha &=& -(\kappa + \Delta\omega_c) \alpha - \rmi g \sum_{j=1}^N  \beta_j - \rmi \frac{\Omega}{2} \,, \\\label{dbeta}
  \frac{\rmd}{\rmd t} \beta_j &=& -\rmi \Delta\omega_x \beta_j + \rmi g \alpha \zeta_j \,, \\\label{dzeta}
  \frac{\rmd}{\rmd t} \zeta_j &=& 2 \rmi g (\alpha^* \beta_j - \alpha \beta_j^*) \,.
\end{eqnarray}
In these equations $\alpha=\langle a\rangle$, $\beta_j=\langle\sigma_-^{(j)}\rangle$, $\zeta_j=\langle\sigma_z^{(j)}\rangle$, $\kappa$ is a cavity loss rate, and $\Delta\omega_y = (\omega_y-\omega_d)$ for $y=c,x$.
Equations~\eqref{dbeta} and~\eqref{dzeta} conserve the length of each pseudospin $4 |\beta_j|^2 + \zeta_j^2 = 1$ as well as that of the total pseudospin $4 |\sum_{j=1}^N\beta_j|^2 + (\sum_{j=1}^N\zeta_j)^2 = C$.
The possible values of $C$ are $C=N^2,(N-2)^2,\dots,c$ where $c=0$ for even $N$ and $c=1$ for odd $N$.
We are interested in steady-state solutions to Eqs.~\eqref{dalpha}--\eqref{dzeta}.
Throughout this work, we assume resonance $\omega_0=\omega_c=\omega_x=\omega_d$ such that $\Delta\omega_c=\Delta\omega_x=0$.
Then, Eq.\eqref{dbeta} requires either $\alpha=0$ or $\zeta_j=0$ for each $j=1,\dots,N$.
Assuming $\alpha=0$, we find
\begin{equation}
  \sum_{j=1}^N \beta_j = -\frac{g}{\kappa} \,, \qquad
  \sum_{j=1}^N \zeta_j = \sqrt{C - (\Omega / g)^2} \,.
\end{equation}
At finite $\alpha$, the phase transition to $\zeta_j=0$ occurs when all $\zeta_j$ become imaginary.
For $N=1$ we have $C=1$ such that the critical point $\Omega=g$ is recovered~\cite{sc88, ac91, agc92, car15}.
In the case $N=2$, the states with $C=4$ (triplet) show a phase transition at $\Omega=2g$, while the states with $C=0$ (singlet) are in the bimodal phase for every finite driving amplitude $\Omega>0$.
The impact of the counter-rotating interaction terms, that couple states with different $C$, will be investigated in the next section.

\paragraph*{Markovian open system dynamics.}
In our approach, dissipation arises from the coupling of the laser-driven Dicke system to the environment.
For a bosonic environment the interaction usually takes the form
\begin{equation}
  H_I = -\rmi X \sum_\alpha \lambda_\alpha (b_\alpha - b_\alpha^\dagger) \,,
\end{equation}
where $X$ is a Hermitian system operator and $b_\alpha$ are annihilation operators of environment photons at frequencies $\omega_\alpha$ with coupling constants $\lambda_\alpha$.
We choose $X=-\rmi(a-a^\dagger)$ for the interaction of the cavity and $X=-\rmi(\sigma_-^{(j)}-\sigma_+^{(j)})$ for the interaction of the $j$th emitter with the environment.

In the limit of weak system-environment coupling the open system dynamics can be treated in the Markovian approximation~\cite{car99, bp02, wei12, aga13}.
The resulting master equation for the system density operator $\rho_S(t)$ becomes
\begin{equation}
  \frac{\rmd}{\rmd t} \rho_S(t) = \mathcal{L}(t) \rho_S(t) \,,
\end{equation}
where $\mathcal{L}(t)$ for $t\geq0$ is the generator of a quantum dynamical semigroup~\cite{lin76, gks76}.
The solution of the master equation requires the choice of a computational basis.
Note that the use of uncoupled emitter and cavity states, yielding the quantum optical master equation with energy-independent dissipative constants, is limited to situations where the intra-system coupling is much smaller than the natural system frequencies~\cite{car99,bgb11,paf15}.
For the Dicke model~\eqref{HD} with zero driving ($\Omega=0$), the master equation expressed in the photon-dressed emitter states includes the regime of strong emitter-cavity interaction~\cite{alz06, sb08, bgb11, rsh13, paf15}.
If the driving is finite ($\Omega\neq0$), the coefficients of the master equation become time-independent in the basis of Floquet states~\cite{bbgssw91, gh98, paf17}.
For the matrix elements $\rho_{m,n}(t)=\langle\psi_m(t)|\rho_S(t)|\psi_n(t)\rangle$ of the system density matrix we obtain the two equations of motion~\cite{paf17}
\begin{eqnarray}\label{master-nn}
  \frac{\rmd}{\rmd t} \rho_{n, n}(t) &=& \sum_{k, \nu} \chi(\omega_{k, n, \nu}) |X_{n, k, \nu}|^2 \rho_{k, k}(t) \nonumber\\
    && - \sum_{k, \nu} \chi(\omega_{n, k, \nu}) |X_{k, n, \nu}|^2 \rho_{n, n}(t) \,, \\\label{master-mn}
  \frac{\rmd}{\rmd t} \rho_{m, n}(t) &=& -Z_{m, n} \rho_{m, n}(t) \qquad (m \neq n) \,.
\end{eqnarray}
In these equations, $\omega_{m,n,\nu}=\epsilon_m-\epsilon_n+\nu\omega_d$ is a transition energy with $\nu$ counting the Fourier modes of the periodic states
\begin{equation}
  | \phi_n(t) \rangle = \sum_{\nu=-\infty}^\infty \rme^{-\rmi \nu \omega_d t} | \tilde{\phi}_n(\nu) \rangle \,.
\end{equation}
Accordingly, the transition matrix elements read
\begin{equation}
  X_{m, n, \nu} = \sum_\mu \langle \tilde{\phi}_m(\mu - \nu) | X | \tilde{\phi}_n(\mu) \rangle \,,
\end{equation}
and the coefficients of the non-diagonal equations of motion~\eqref{master-mn} are
\begin{eqnarray}
  Z_{m, n} &=& \frac{1}{2} \sum_{k, \nu} \big[ \chi(\omega_{m, k, \nu}) + \rmi \xi(\omega_{m, k, \nu}) \big] |X_{k, m, \nu}|^2 \nonumber\\
    && + \frac{1}{2} \sum_{k, \nu} \big[ \chi(\omega_{n, k, \nu}) + \rmi \xi(\omega_{n, k, \nu}) \big] |X_{k, n, \nu}|^2 \nonumber\\
    && - \sum_\nu \chi(\nu \omega_d) X_{m, m, \nu} X_{n, n, \nu}^* \,.
\end{eqnarray}
The functions $\chi(\omega)$ and $\xi(\omega)$ are even and odd Fourier transforms of the reservoir correlation function for a thermal environment at temperature $T$, and are given by
\begin{equation}
  \chi(\omega) = \begin{cases} \gamma(\omega) [n(\omega, T) + 1] & \text{if}~\omega > 0 \\ \gamma(-\omega) n(-\omega, T) & \text{if}~\omega < 0 \end{cases}
\end{equation}
and
\begin{equation}
  \xi(\omega) = \begin{cases} \Re \Gamma(\omega + \rmi 0^+) [n(\omega, T) + 1] & \text{if}~\omega > 0 \\ \Re \Gamma(-\omega + \rmi 0^+) n(-\omega, T) & \text{if}~\omega < 0 \end{cases} \,,
\end{equation}
respectively.
In these equations, $\gamma(\omega)$ is the environment spectral function, $\Gamma(\omega)$ is its analytic continuation into the upper half plane, and $n(\omega,T)=(\rme^{\omega/T}-1)^{-1}$ is the thermal distribution function.
Here, we assume the same Ohmic spectral function $\gamma(\omega)=\gamma\omega/\omega_0$ for both, cavity and emitter environments.
The respective environment temperatures are also identical.

\begin{figure}
\includegraphics[width=\linewidth]{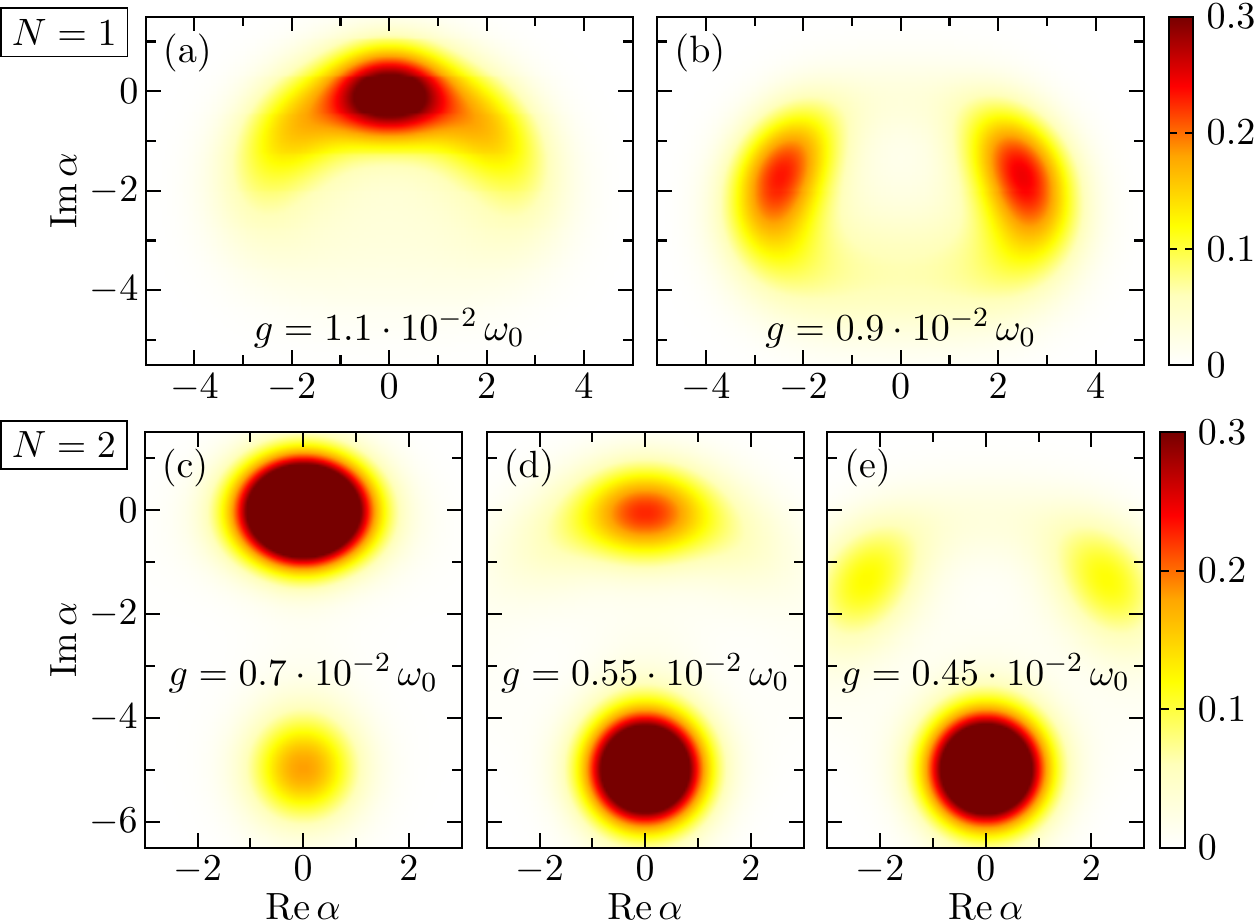}
\caption{\label{fig:bistab}Husimi function of the stationary field state in the long-time limit for $N=1$ (a) and (b), and $N=2$ (c)--(e).
  The driving amplitude $\Omega=10^{-2}\,\omega_0$, the temperature $T=0$, and $\gamma=10^{-3}\,\omega_0$.
  Note that the same dark color is assigned to $Q(\alpha)\geq0.3$ in the density plots.}
\end{figure}

As a first application of this formalism, we consider the (unnormalized) Husimi function
\begin{equation}\label{Q}
  Q(\alpha) = \langle \alpha | \rho_c^{ss} | \alpha \rangle
\end{equation}
of the stationary field state $\rho_c^{ss}$ in the long-time limit, that is commonly used to visualize the dissipative quantum phase transition~\cite{sc88, ac91, car15}.
In Eq.~\eqref{Q}, $|\alpha\rangle$ is a coherent state with the complex number $\alpha\in\mathbb{C}$.
Evaluation of the Husimi function allows us to determine whether the phase transition still exists when the counter-rotating terms are taken into account, as is required in the regimes of strong emitter-cavity coupling and high driving amplitudes.
Because the state in the long-time limit keeps oscillating with the driving frequency, we take $\rho_c^{ss}=\lim_{n\to\infty}\mathop{Tr}_x\rho_S(nT_d)$ at the beginning of a modulation period with $\mathop{Tr}_x$ denoting the partial trace over the emitter degrees of freedom.
Figure~\ref{fig:bistab} demonstrates a clear bimodality for one emitter when the critical point at $\Omega = g$ is passed.
For two emitters, the Husimi function in Figs.~\ref{fig:bistab}(c)--\ref{fig:bistab}(e) exhibits a more complex behavior.
For parameters well below the critical point, the Husimi function in panel (c) indicates a mixture of two states.
These states correspond to the two solutions of Eqs.~\eqref{dalpha}--\eqref{dzeta}, i.\,e., to the triplet ($C=4$) at $\alpha=0$ and the singlet ($C=0$) at $\alpha=-5\rmi$.
Lowering the emitter-cavity coupling strength at fixed driving amplitude, the weights for the two states of the mixture change, see panel (d) of Fig.~\ref{fig:bistab}.
This changing is caused by the coupling of singlet and triplet states through the counter-rotating interaction terms.
Above the critical point $\Omega=2g$, the triplet state at $\alpha=0$ shows the expected bimodality.
We thus conclude that the phase transition occurs also when going beyond the rotating-wave approximation.

\paragraph*{Non-Markovian emitter dynamics.}
A general measure for the degree of non-Markovian behavior in open quantum systems was introduced in Ref.~\cite{blp09}.
It is based on the change of the trace distance,
\begin{equation}
  D[\rho_1(t), \rho_2(t)] = \frac{1}{2} \text{tr}|\rho_1(t) - \rho_2(t)| \,,
\end{equation}
of two time-dependent states $\rho_{1,2}(t)=\Phi(t,0)\rho_{1,2}(0)$.
For a given quantum process $\Phi(t,0)$, the non-Markovianity is defined as
\begin{equation}\label{nonMarkov}
  \mathcal{N} = \max_{\rho_{1,2}(0)} \int_{\sigma > 0} \sigma[t, \rho_{1,2}(0)] \, \rmd t \,,
\end{equation}
where
\begin{equation}
  \sigma[t, \rho_{1,2}(0)] = \frac{\rmd}{\rmd t} D[\rho_1(t), \rho_2(t)]
\end{equation}
is the change of the trace distance between $\rho_1(t)$ and $\rho_2(t)$.
The quantity $\mathcal{N}$ in Eq.~\eqref{nonMarkov} measures the increase in distinguishability of the two quantum states.
If one interprets the loss of distinguishability in a Markovian process as a flow of information from the system to the environment, the increase in distinguishability during non-Markovian dynamics belongs to the inverse process, where information flows from the environment back to the system~\cite{blp09}.
The use of the non-Markovianity $\mathcal{N}$ is not limited to theoretical studies, but can be tested experimentally without knowledge about the structure of the environment or the interaction between system and environment~\cite{lpb10}.
It was used to analyze a possible speedup of open system evolution~\cite{zxf16}.

The quantum process defined by Eqs.~\eqref{master-nn} and~\eqref{master-mn} belongs to a dynamical semigroup and has the property, that the distance between any pair of quantum states monotonically decreases, i.\,e., it is a Markovian process.
Non-Markovianity manifests itself if the system dynamics of emitters and cavity is projected onto a subspace.
Here, we are interested in the non-Markovianity of the emitter dynamics obtained by performing the partial trace of the total system density matrix over the cavity degrees of freedom.
In this case, the non-Markovianity $\mathcal{N}$ quantifies the flow of information from the cavity mode to the emitters which can be interpreted as a negative effective dissipation for some intervals of time~\cite{blp09}.

In the numerical computations we use $\gamma=0.01\,\omega_0$ and $T=0.05\,\omega_0$.
Additionally, a maximum number of 50 cavity photons and 110 Fourier modes are used for evaluating the Floquet states and we carefully checked that the results do not change if these numbers are increased.

\begin{figure}
  \includegraphics[width=\linewidth]{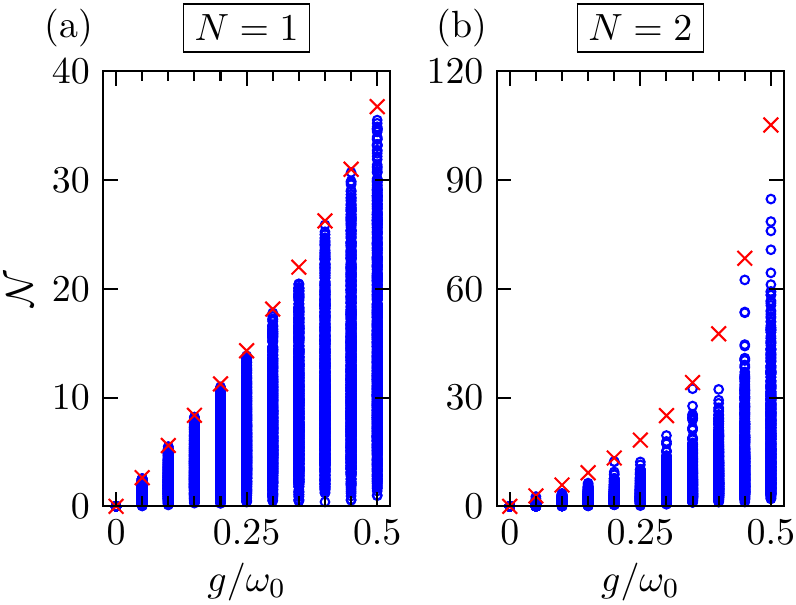}
  \caption{\label{fig:nonMarkov_nd}Non-Markovianity $\mathcal{N}$ without drive ($\Omega=0$) for (a) $N=1$ and (b) $N=2$ as functions of the emitter-cavity coupling $g$.
    Blue circles depict randomly drawn pairs of initial states and red crosses mark the initial pair given by Eq.~\eqref{preparation1}.}
\end{figure}

Let us first discuss the results for zero driving and a single emitter ($N=1$), see Fig.~\ref{fig:nonMarkov_nd}(a).
They are obtained through numerical integration of Eqs.~\eqref{master-nn} and~\eqref{master-mn} and a subsequent projection onto the emitter degrees of freedom.
The maximum over pairs of initial states $\rho_{1,2}(0)$ in Eq.~\eqref{nonMarkov} is computed by preparation and propagation of a sufficiently large sample of random initial states.
Our data provide strong evidence, that the maximum is always attained for the initial states
\begin{equation}\label{preparation1}
  \rho_{1,2}(0) \equiv \rho_\pm(0) = \frac{1}{2} \big( | + \rangle \pm | - \rangle \big) \big( \langle + | \pm \langle - | \big) \,.
\end{equation}
This result is in contrast to the finding reported in Ref.~\cite{lpb10} for a two-level atom coupled to a vacuum reservoir, where the maximum is attained for the pair $|-\rangle\langle-|$ and $\rho_+(0)$.
Nevertheless, as shown in Refs.~\cite{xyf10, blpv16}, this finding is wrong---the states~\eqref{preparation1} yield the maximum non-Markovianity also in the model discussed in Ref.~\cite{lpb10}.
In our case, the strong coupling to the additional cavity mode and the small but finite environment temperature leads to the increased impact of the coherences to the overall non-Markovianity.
Hence, the largest $\mathcal{N}$ is attained for states, which have a maximum initial trace distance and the highest possible non-diagonal elements.
This fact can be proven mathematically for any open quantum system with a linear dynamical map $\Phi(t,0)$~\cite{wklpb12}.
This prove remains valid here, because also the partial trace over cavity degrees of freedom is linear.

\begin{figure}
  \includegraphics[width=\linewidth]{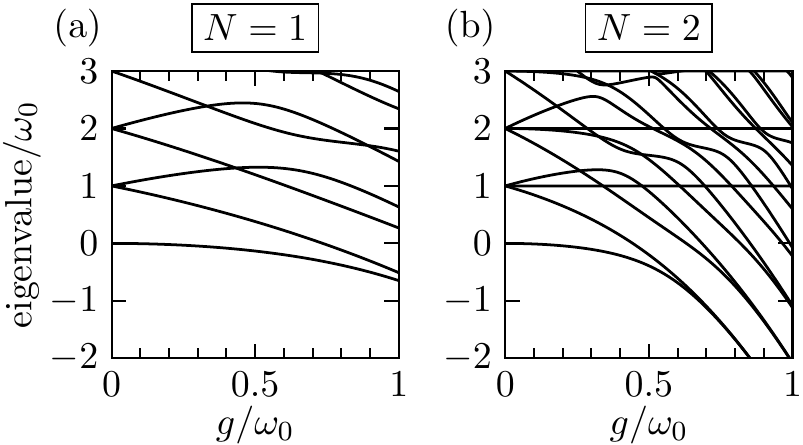}
  \caption{\label{fig:disp}Eigenvalues of $H_D$ in Eq.~\eqref{HD} for (a) $N=1$ and (b) $N=2$ as functions of the emitter-cavity coupling $g$.}
\end{figure}

The dynamics of the emitter is completely Markovian if the emitter-cavity coupling is zero, that is, when the system dynamics factorizes into independent emitter and cavity evolutions.
For finite emitter-cavity coupling strength $g\neq0$, the non-Markovianity becomes finite and increases almost linearly with $g$.
The linear increase is explained by the dispersions of the dressed emitter-cavity states starting linearly for small coupling, see Fig.~\ref{fig:disp}.
Deviations from the linear behavior are visible for $g\gtrsim0.4\,\omega_0$.
This boundary should be significantly smaller if the number of emitters is increased.

Figure~\ref{fig:nonMarkov_nd}(b) depicts the results for two emitters with $\Omega=0$.
The maximum in Eq.~\eqref{nonMarkov} is attained for the states
\begin{multline}\label{preparation2}
  \rho_{1,2}(0) \equiv \rho_\pm(0) = \frac{1}{4} \big( | +, + \rangle \pm | +, - \rangle \pm | -, + \rangle + | -, - \rangle \big) \\
    \times \big( \langle +, + | \pm \langle +, - | \pm \langle -, + | + \langle -, - | \big) \,.
\end{multline}
Again, these states have the maximum initial trace distance of $D[\rho_1(0),\rho_2(0)]=1$ and the highest possible coherence.
The linear g-dependence of the non-Markovianity $\mathcal{N}$ in Fig.~\ref{fig:nonMarkov_nd}(a) for a single emitter is changed to a superlinear one in Fig.~\ref{fig:nonMarkov_nd}(b) if the number of emitters is increased.
Again, this behavior can be traced back to the nonlinear dispersions of the dressed emitter-cavity states in Fig.~\ref{fig:disp}.
Deviations from the linear relation occur for $g\gtrsim0.2\,\omega_0$.

\begin{figure}
  \includegraphics[width=\linewidth]{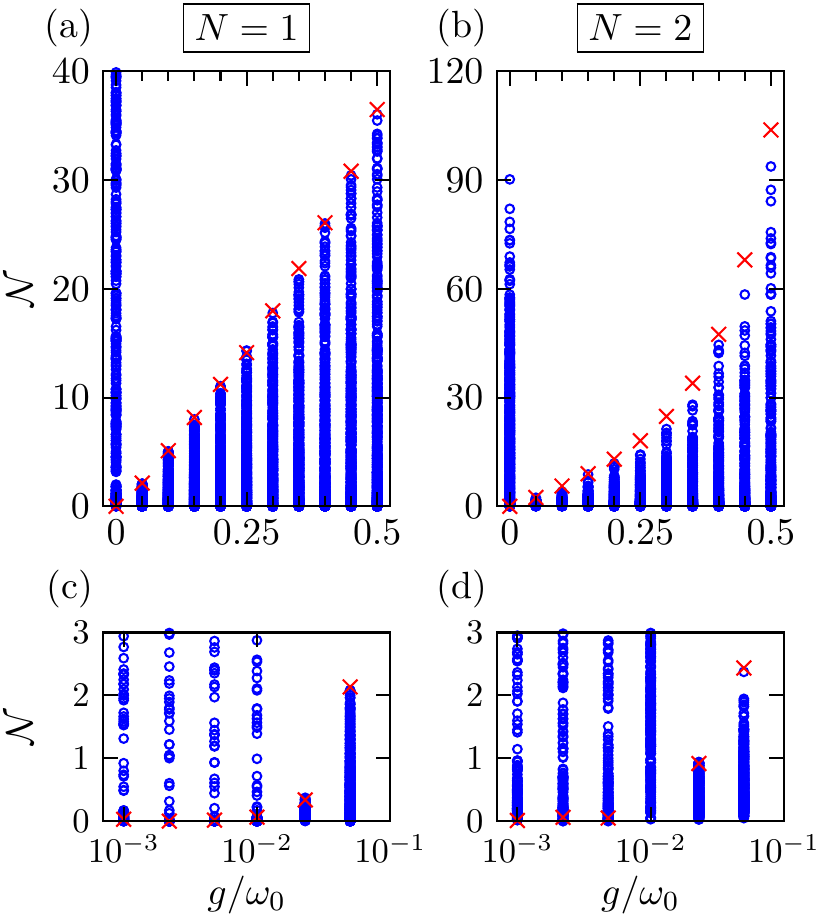}
  \caption{\label{fig:nonMarkov_wd}Non-Markovianity $\mathcal{N}$ with drive ($\Omega=10^{-2}\,\omega_0$) for (a), (c) $N=1$ and (b), (d) $N=2$ as functions of the emitter-cavity coupling $g$.
    Panels (c) and (d) give a zoom into the weak coupling regime.
    Blue circles depict randomly drawn pairs of initial states and red crosses mark the initial pair given by Eq.~\eqref{preparation1}.}
\end{figure}

For finite driving the numerical results for the non-Markovianity $\mathcal{N}$ of $N=1$ and $N=2$ emitters are given in Fig.~\ref{fig:nonMarkov_wd}.
Obviously, the results for $g>\Omega$ remain unchanged when they are compared to the situation without external laser driving in Fig.~\ref{fig:nonMarkov_nd}.
This is particularly interesting because both, the stationary state in the long-time limit and the whole system dynamics towards it, are changed with finite driving amplitude.
Nevertheless, the amount of information flowing between the emitters and the cavity photons is independent of $\Omega$, showing that it depends only on the (unchanged) interaction between emitters and cavity, but not on the system-environment interaction that is responsible for dissipation.

\begin{figure}
  \includegraphics[width=\linewidth]{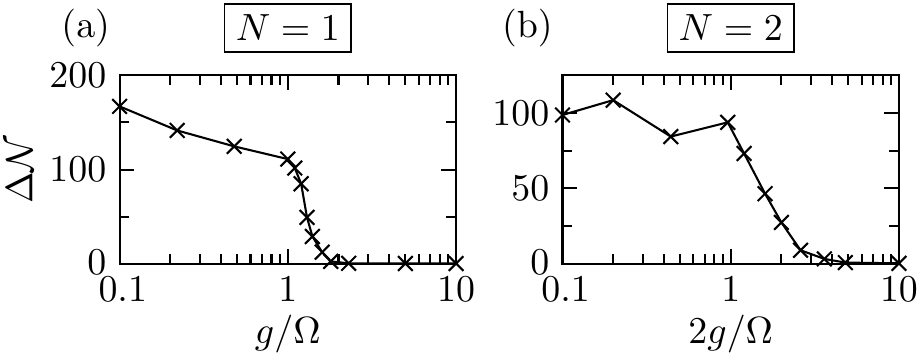}
  \caption{\label{fig:deltaN}Non-Markovianity change $\Delta\mathcal{N}$ as a function of the emitter-cavity coupling $g$ at fixed $\Omega=10^{-2}\,\omega_0$ for (a) $N=1$ and (b) $N=2$.}
\end{figure}

Comparing Fig.~\ref{fig:nonMarkov_wd} with Fig.~\ref{fig:nonMarkov_nd} we observe an enhanced non-Markovianity in the regime of strong driving $\Omega>g$.
The origin of this enhancement is the quantum dissipative phase-transition, where the coupled emitter and cavity modes undergo spontaneous dressed state polarization~\cite{sc88, ac91, car15}.
Above the critical driving strength, the intracavity field shows a bimodality in phase which is caused by the intensity-dependent detuning between the driving field and the transition frequencies of the composite emitter-cavity system (without drive).
The oscillations between these bimodal states, which are induced by the environmental coupling, are responsible for the enhanced non-Markovianity.
Usually, the bimodality is monitored by the Husimi function, see Fig.~\ref{fig:bistab}.
Here, the non-Markovianity $\mathcal{N}$ directly measures the information flow between emitters and cavity that is related to the spontaneous transitions between dressed states.
Thus, the non-Markovianity change with increasing driving strength,
\begin{equation}
  \Delta \mathcal{N} = |\mathcal{N}_{\Omega>0} - \mathcal{N}_{\Omega=0}| \,,
\end{equation}
could serve as an indicator of the quantum dissipative phase transition.
As shown in Fig.~\ref{fig:deltaN}, $\Delta\mathcal{N}$ is zero above and finite below the critical point at $\Omega=Ng$.
The small but finite temperature $T = 0.05 \, \omega_0$ used in the numerical calculation leads to a smooth transition that appears at a slightly enhanced emitter-cavity coupling strength.
Hence, a finite temperature stabilizes the normal phase below the critical point.

\paragraph*{Conclusions.}
Our analysis of the non-Markovian dynamics of few emitters coupled to a laser-driven cavity requires the use of Floquet states as the computational basis.
In this way, the regimes of strong emitter-cavity coupling as well as high field amplitudes of the driving laser can be studied beyond the rotating-wave approximation.
This is particularly important in view of recent experimental advances on strongly driven two-level systems interacting with a bosonic environment~\cite{rs17, xzsyjxlcy17}.
While the system evolution, for weak coupling to the environment, is described with a Markovian master equation, the projection onto the emitter degrees of freedom yields a non-Markovian dynamical process.
Evaluation of the non-Markovianity thus allows for a quantification of the information flow between emitters and cavity photons.

The numerical data for the non-Markovianity provide strong evidence that its maximum is always attained for the pair of pure initial states which have the highest possible trace distance and which have the largest non-diagonal elements.
This result is in accordance with the mathematical prove presented in Ref.~\cite{wklpb12} showing that the states have to be orthogonal and must lie on the boundary of the space of physical states.
Nevertheless, these two properties are not sufficient to determine the exact states for more than one emitter.
In this sense, our result, which is explained by the high impact of the coherences in the regimes of strong emitter-cavity coupling and high driving amplitudes, goes beyond the scope of this prove.

Note that the behavior of the non-Markovianity with increasing emitter-cavity coupling strength strongly depends on the number of emitters in the cavity.
We find a linear dependence for a single emitter and a superlinear one for two emitters.
Interestingly, this behavior is independent of the laser driving strength if the emitter-cavity coupling is large enough.
Specifically, we observe an increased non-Markovianity above the dissipative quantum phase transition occurring if the laser driving strength is larger than the emitter-cavity coupling strength.
Thus, the change of the non-Markovianity with the laser-driving strength can be taken as an indicator of the phase transition.
Theoretically, the advantage of such an indicator is, that a search for a bimodal Husimi function is not needed because it is a single real-valued number.
Besides, a lower bound for the non-Markovianity can already be obtained from a measurement of populations such that a complete tomographic state reconstruction is not necessary~\cite{xyf10}.
Therefore, the non-Markovianity change will also be feasible for future experimental implementation.

\begin{acknowledgments}

This work was supported by Deutsche Forschungsgemeinschaft through SFB 652 and SFB/TRR 24.

\end{acknowledgments}


%

\end{document}